\documentclass{article}
\usepackage{dcase2025_techrep,amsmath,graphicx,url,times,booktabs, tabularx,gensymb}
\usepackage{amssymb}

\title{Stereo sound event localization and detection based on PSELDnet pretraining and BiMamba sequence modeling}

\twoauthors
  {Wenmiao Gao}
    {Denmark Technical University\\
    2800 Kgs. Lyngby Denmark \\
     wenmiaogao@163.com}
  {Yang Xiao}
    {  The University of Melbourne\\
     Melbourne, Australia\\
     yxiao9550@student.unimelb.edu.au\
    }

\begin{document}

\ninept
\maketitle

\begin{sloppy}

\begin{abstract}
Pre-training methods have achieved significant performance improvements in sound event localization and detection (SELD) tasks, but existing Transformer-based models suffer from high computational complexity. In this work, we propose a stereo sound event localization and detection system based on pre-trained PSELDnet and bidirectional Mamba sequence modeling. We replace the Conformer module with a BiMamba module and introduce asymmetric convolutions to more effectively model the spatiotemporal relationships between time and frequency dimensions. Experimental results demonstrate that the proposed method achieves significantly better performance than the baseline and the original PSELDnet with Conformer decoder architecture on the DCASE2025 Task 3 development dataset, while also reducing computational complexity. These findings highlight the effectiveness of the BiMamba architecture in addressing the challenges of the SELD task.
\end{abstract}


\section{Introduction}
\label{sec:intro}

The objective of the sound event localization and detection (SELD) task is to identify sound events from predefined target classes, track their temporal dynamics, and estimate their respective spatial trajectories where they are active\cite{Adavanne2018Jun}.This technology plays a vital role in various real-world applications, such as robotic auditory sensing, human-compute interaction, and immersive audio experiences.

Since 2019, the Sound Event Localization and Detection (SELD) task has been featured as an annual challenge in the DCASE competition, attracting an increasing number of researchers and driving dramatic progress in the field. From 2019 to 2021, the datasets consisted primarily of clean sound events convolved with spatial room impulse responses (SRIRs) recorded in various rooms. In 2022, real recorded audio data was introduced for the first time as part of the task dataset. In 2023, the task was further expanded from an audio-only track to an audio-visual track, with 360-degree panoramic videos and corresponding audio provided for the dataset. In 2024, a new distance estimation subtask was introduced. This year, the original FOA audio and 360-degree video have been converted into stereo audio and perspective video, simulating the scenario of regular media content.

Deep learning methods have significantly advanced the field by greatly improving the performance of both sound event classification (SEC) \cite{Heittola2013Dec,7096611,fmsg_dcase2024,wilddesed,ucil,xiao2024mixstyle} and direction of arrival (DOA) estimation \cite{Schmidt1986Mar,xiao2025wheresvoicecomingcontinual} compared to traditional machine learning and signal processing approaches. Adavanne et al. \cite{Adavanne2018Jun} proposed the first end-to-end SELDnet, enabling simultaneous sound event detection and localization. However, this approach could not address the overlap problem, where multiple sound sources of the same class occur simultaneously at different locations. To tackle this issue, the track-wise format and permutation invariant training were introduced in EINv2 \cite{Cao}, which employs two parallel branches for SED and DOA tasks, as well as an additional parameter-sharing branch. Nevertheless, such multi-branch SELD models require balancing the loss functions of both tasks during training, which increases system complexity and computational cost. To address these limitations, the class-wise ACCDOA \cite{Shimada2020Oct} output format was proposed, which integrates SED and localization tasks into a single Cartesian vector, allowing the SELD task to be solved as a single-target problem. Furthermore, the Auxiliary Duplicated Permutation Invariant Training (ADPIT) method was introduced to effectively handle the homogenous overlap issue, the MultiACCDOA  \cite{Shimada2021Oct} is currently recognized as the baseline output format.

Previous studies have shown that increasing the amount of training data can significantly enhance the performance of SELD systems. This is also why data augmentation techniques such as Audio Channel Swapping (ACS) have become highly popular, as swapping audio channels can effectively augment the amount of real data. However, it is challenging to collect or simulate large-scale spatial audio data of specific types in real-world scenarios. Therefore, fine-tuning pre-trained SELD models with a limited amount of data has become a practical solution to address the issue of insufficient training samples. Hu et al. \cite{Hu2024Nov} proposed PSELDnet, which is built upon state-of-the-art pre-trained architectures for sound event classification, such as PANNs \cite{Kong2019Dec}, PaSST \cite{Koutini2021Oct}, and HTS-AT \cite{Chen2022Feb}, and is trained on large-scale synthetic datasets. Experimental results demonstrate that PSELDnet dramatically outperforms the baseline across multiple downstream datasets.

In this year’s challenge, the STARSS23 \cite{Shimada2023Jun} dataset has been adapted to a fixed-perspective setting, with FOA audio converted to mid-side (M/S) stereo format. This adaptation makes it considerably more difficult to generate simulated data, as it requires the simultaneous generation of both spatially consistent audio and video data, which is technically challenging and resource-intensive. As a result, fine-tuning the PSELDnet pre-trained model directly on the stereo dataset becomes a natural and effective solution for this task.

In addition to pretrained models, the Mamba \cite{Gu2023Dec} architecture successfully combines CNN's local feature extraction capabilities with Transformer's global modeling advantages through selective state space (SSM), while maintaining linear computational complexity O(n), achieving SOTA performance across multiple audio and speech domains. In the field of speech separation, Li et al. \cite{Li2024Apr} replaced BLSTM in the TF-GridNet architecture with BiMamba, achieving SOTA results on multiple datasets. Zhang et al. \cite{Zhang2024May} employed BiMamba as a replacement for MHSA, demonstrating excellent performance in speech enhancement tasks. Mu et al. \cite{Mu2024Aug} replaced the Conformer module with BiMamba based on the EINV2 framework, surpassing EINV2's performance in multi-task output SELD systems. Additionally, Liu \cite{Liu2024Nov} proposed ND-BiMamba2 based on Mamba2 \cite{Dao2024May}, which can effectively process 2D data.

In this report, we focus on the audio-only track and use the official development dataset, which is derived from the STARSS23 dataset and converted to stereo format through Mid-side conversion. To meet the input requirements of the pretrained PSELDnet, we regenerate pseudo FOA data from stereo signals. During training, we employ audio channel swap (ACS) (\cite{Wang2021Jan}) for data augmentation. We comprehensively fine-tune the pretrained PSELDnet models (including multi ACCDOA HTS-AT model and multi ACCDOA CNN14-Conformer model) for the SELD task, replacing the Conformer architecture with BiMamba architecture and introducing asymmetric convolutions to reduce computational complexity and decouple audio time-frequency domain features. Experimental results demonstrate that the BiMamba architecture performs excellently in SELD tasks, significantly improving system performance, while the general PSELDnet can be effectively transferred to stereo SELD tasks, greatly surpassing the baseline performance.
\section{THEORY}
\label{sec:theory}
The Mamba architecture, based on the S4 (Structured State Space Sequence) model \cite{Gu2021Oct,xiao2025tfmambatimefrequencynetworksound}, effectively combines the advantages of CNN and RNN, allowing the use of CNN's parallel computation benefits during training and RNN's temporal modeling capabilities during inference. The introduction of state-selection mechanisms allows the model to selectively attend to or ignore specific parts of the input sequence, which is crucial for accurate discrimination of overlapping audio events.

  Specifically, inspired by continuous linear time-invariant systems in signal processing and control systems, it transforms the input sequence $x(t) \in \mathbb{R}$ to the output sequence $y(t) \in \mathbb{R}$ using higher dimensional hidden states $h(t) \in \mathbb{R}^{N \times 1}$, which can be written as follows:
\begin{equation}
\begin{aligned}
&    \mathbf{h}'(t) = \mathbf{A}\mathbf{h}(t) + \mathbf{B} x(t) \\
 &   y(t) = \mathbf{C}^T \mathbf{h}'(t) + \mathbf{D}x(t)
    \label{equ: 1}
\end{aligned}
\end{equation}
where $\mathbf{A} \in R^{N\times N}$, $\mathbf{B} \in R^{N \times 1}$,$\mathbf{C} \in R^{N \times 1}$ , and $\mathbf{D}$ represent the state transition matrix, the input projection matrix, the output projection matrix, and the skip connection matrix respectively.

In practice, to handle discrete sequences, discretization of the SSM is necessary. Using the Zero Order Holding method, we introduce a time step $\Delta$ to sample the continuous matrices $\mathbf{A}$ and $\mathbf{B}$, obtaining discrete representations $\bar{\mathbf{A}}$ and $\bar{\mathbf{B}}$, as follows:

\begin{equation}
\begin{aligned}
    & \bar{\mathbf{A}} = \exp(\Delta \mathbf{A})\\
    &\bar{\mathbf{B}} = (\Delta \mathbf{A})^{-1}(\exp \Delta \mathbf{A} - \mathbf{I}) \cdot \Delta \mathbf{B}
    \label{equ:2}
\end{aligned}
\end{equation}
and the discretized structured SSM are as follows:
\begin{equation}
\begin{aligned}
    & \mathbf{h}_k = \bar{\mathbf{A}}\mathbf{h}_{k-1} + \bar{\mathbf{B}}x_k
    \\
    & y_k = \mathbf{C}^T \mathbf{h}_k
    \label{equ: 4}
 \end{aligned}
\end{equation}

The above discretized parameters vary over time using a selective state space modeling (SSM) approach, similar to the gating mechanism in RNNs, enabling the model to selectively attend to or ignore input features at each time step, thereby enhancing the model's information processing capability.

\section{PROPOSED METHOD}

\label{sec:method}

\subsection{Feature Extraction}
Since PSELDNet operates on FOA-format audio data by concatenating 4-channel log-mel spectrograms with 3-channel intensity vectors, to meet the input feature requirements of the pre-trained network, we convert the stereo left-right ear signals $L(n)$ and $R(n)$ back to the FOA components $W(n)$ and $Y(n)$ according to the ACN/SN3D convention. The remaining components $X(n)$ and $Z(n)$ are set to zero, as formulated in Equation (\ref{equ:3})
\begin{equation}
\begin{aligned}
    W(n) &= \frac{L(n) + R(n)}{2} \\
    Y(n) &= \frac{L(n) - R(n)}{2} \\
    X(n) &= 0\\
    Z(n) &= 0
    \label{equ:3}
\end{aligned}
\end{equation}
These parameters form a four-channel pseudo FOA representation. Subsequently, we extract the corresponding log-mel spectrograms and intensity vectors, which are then concatenated to construct a 7-channel input feature tensor.

\subsection{Data Augmentation}
To mitigate overfitting and enhance the overall system's robustness and generalization capability, we primarily employ Audio Channel Swapping (ACS) to exchange left-right ear channels for generating augmented Direction of Arrival (DOA) labels. In this year's challenge, DOA labels were rotated to a fixed frontal perspective as the reference coordinate system, which introduces front-back ambiguity. To resolve this, the azimuth labels are folded into the range [-90°, 90°] through backward-forward mapping, while elevation angles remain excluded from consideration this year. Through ACS implementation, the development dataset is effectively doubled via geometric symmetry exploitation.

\subsection{Network Architecture}
In this work, we improve upon the PSELDnet pretrained models. PSELDnet primarily consists of two pretrained architectures: CNN14-Conformer hybrid architecture and HTS-AT pure Transformer architecture. Our main innovation lies in replacing the Conformer decoder in the CNN14-Conformer architecture with a BiMamba module combined with asymmetric convolutions as the decoder, forming the CNN14-BiMamba hybrid architecture.

Specifically, the CNN14-BiMamba architecture employs CNN14 as the encoder and BiMamba module combined with asymmetric convolutions as the decoder. The CNN14 backbone network consists of six VGG-style convolutional blocks with 3×3 convolutions, batch normalization, and ReLU activation, enabling hierarchical extraction of fine-grained spectral-temporal features. The BiMamba module, based on the Mamba architecture, overcomes the causal limitations of traditional Mamba through bidirectional processing mechanisms, enabling simultaneous capture of forward and backward temporal dependencies. In the specific implementation, we separately process input features through forward and backward Mamba processing, then concatenate and fuse the hidden representations from both directions, finally compressing the channel dimensions through a linear layer to obtain the output. The introduction of asymmetric convolutions further enhances the model's ability to decouple time-frequency domain features in stereo SELD tasks, effectively improving recognition performance for overlapping audio events through alternating processing of time and frequency dimensions.

Previous work \cite{Zhang2024May} has shown that replacing the MHSA module in Conformer with BiMamba can improve the capture performance of low-abstraction-level spectral information in speech enhancement tasks. Here, we also attempt to apply this method to stereo SELD tasks.

For comparative experiments, we use the following models: First, the original CNN14-Conformer model, which is a CNN-attention hybrid network where CNN14 contains a stack of 6 VGG-style CNN blocks, and the Conformer module contains two feed-forward layers sandwiching multi-head self-attention and convolution modules with residual connections. The CNN blocks extract local fine-grained features, while the Conformer blocks capture local and global contextual dependencies in audio sequences.
Second, the HTS-AT model combines Swin Transformer and token-semantic modules. The Swin Transformer focuses on self-attention within each local window, establishing connections between consecutive layers through shifted window attention mechanisms and building hierarchical feature maps. The token-semantic module employs a convolutional layer as the head layer, converting feature maps to activation maps for timestamp prediction.

Finally, the traditional CRNN baseline model combines the advantages of convolutional neural networks and recurrent neural networks, providing a benchmark reference for our experiments.

In the final stage of the two main models, we apply Tanh activation function to DoA (Direction of Arrival) estimates and ReLU activation function to distance estimates. The Tanh function constrains DoA outputs to the range $[-1,1]$, which can be mapped to azimuth angle range $[-90\degree, 90\degree]$, ensuring predictions remain within the valid field of view. The ReLU function enforces non-negative distance estimates, which is physically meaningful as distances cannot be negative. This design choice aligns model outputs with the actual constraints of the task, as we primarily focus on azimuth angles within the visible light range and require all distance values to be non-negative.

\section{Training}
\label{sec:training}
The proposed system is trained on the development dataset (stereo format) of STARSS23\cite{Shimada2025Apr}. The audio is resampled to 24 kHz, and 64 mel filters are used for feature extraction. The short-time Fourier transform (STFT) is computed with a hop length of 20 ms (480 samples) and a window length of 40 ms. Each input feature consists of 250 frames, corresponding to 50 label frames.

For training, the random seed is set to be 42. Due to different parameter counts and memory requirements across models, we employ different batch sizes, learning rates, and weight decay values. To simplify model names, we uniformly omit the prefix "CNN14". The training configurations for each model are as follows: HTSAT and Conformer models use a learning rate of 1e-4, weight decay of 1e-4, and batch size of 256; ConBiMamba uses 1e-4, 5e-6, and 32 respectively; BiMamba uses 1e-4, 5e-6, and 128; and BiMambaAC uses 3e-5, 5e-6, and 32. All models are trained for 120 epochs using the Adam optimizer with a ReduceLROnPlateau scheduler (mode='max', reduction factor=0.5, patience=5). Following the baseline protocol, we evaluate our models on the test split of the DCASE2025 Task 3 development set and select the checkpoint with the best validation location-dependent F1-Score as the final model.

Full fine-tuning of all model parameters is adopted due to the substantial differences between the input features and those used in the pre-trained models, as well as the introduction of a new task. This allows the model to better adapt to the new data distribution and task-specific objectives, which may not be achievable through partial fine-tuning.
\section{Results}
\label{sec:Results}

Table \ref{tab:performance} presents a quantitative comparison between the baseline system and various pre-trained models on the development set. The metrics reported include the number of model parameters, the location-dependent F1-score ($F_{20^\circ}$, higher is better), the Direction of Arrival Error (DOAE, lower is better), and the Relative Distance Error (RDE, lower is better).

The baseline system contains only 0.7 million parameters and achieves an $F_{20^\circ}$ of 22.8\%, with a DOAE of 24.5° and an RDE of 41\%. Among the pre-trained models, HTS-AT (28M parameters) shows significant improvements with an $F_{20^\circ}$ of 35.1\% and a DOAE of 15.8° while maintaining the lowest RDE of 30\%. The Conformer model (210M parameters) achieves comparable performance with an $F_{20^\circ}$ of 38.2\% and a DOAE of 16.6°.

The BiMamba-based models exhibit different performance characteristics. ConBiMamba (338M parameters), while performing well in speech enhancement tasks, shows mediocre performance in stereo SELD tasks with a large parameter count and significant distance localization error (RDE of 53\%). BiMamba (178M parameters) achieves an $F_{20^\circ}$ of 36.2\% with a DOAE of 16.6°. Notably, BiMambaAC (76M parameters, 4.63G MACs) achieves the best overall performance with an $F_{20^\circ}$ of 39.6\%, a DOAE of 15.8°, and an RDE of 33\%. Compared to BiMamba, the incorporation of asymmetric convolution significantly improves model performance while reducing parameter count, achieving superior results across all metrics with only 76M parameters and 4.63G MACs.

\begin{table}[htbp]
\centering

\caption{Comparison of different model architectures on the development set}
\label{tab:performance}
\resizebox{\linewidth}{!}{
\begin{tabular}{lccccc}
\toprule
Model & Params & MACs & $F_{20^\circ} \uparrow$ & $DOAE \downarrow$ & $RDE \downarrow$ \\
\midrule
BiMambaAC (s1) & $76M$ & $4.63G$ & \textbf{39.6\%} & \textbf{15.8\degree} & 33\% \\
BiMamba (s2)& $178M$ & $7.57G$ & $36.2\%$ & $16.6\degree$ & $33\%$ \\
Conformer (s3) & $210M$ & $4.69G$ & $38.2\%$ & $15.9\degree$ & $33\%$ \\
HTS-AT (s4) & $28M$ & $2.88G$ & $35.1\%$ & $16.5\degree$ & $\textbf{30\%}$ \\
ConBiMamba & $338M$ & $7.98G$ & $36.2\%$ & $16.9\degree$ & $53\%$ \\
Baseline & $0.7M$ & $57M$ & $22.8\%$ & $24.5\degree$ & $41\%$ \\
\bottomrule
\end{tabular}}
\end{table}

\section{Conclusion}
\label{sec:conclustion}
In this work, we propose a novel approach for stereo sound event localization and detection (SELD) by combining PSELDnet pretraining with BiMamba sequence modeling. Our experimental results demonstrate several key findings:First, the integration of pre-trained models significantly improves SELD performance compared to the baseline system. The HTS-AT model achieves substantial improvements in both F1-score and DOAE, while the Conformer model maintains competitive performance.
Second, BiMamba-based architectures show superior performance in stereo SELD tasks. The BiMamba model achieves improved F1-score and DOAE compared to traditional architectures, demonstrating the effectiveness of state space modeling for audio sequence processing.
Most importantly, our proposed BiMambaAC model, incorporating asymmetric convolution with BiMamba, achieves the best overall performance across all metrics while maintaining a relatively compact parameter count. This demonstrates that the combination of asymmetric convolution and BiMamba architecture not only enhances model performance but also reduces computational complexity compared to larger models like ConBiMamba.
The results validate the effectiveness of our approach in leveraging pre-trained models and advanced sequence modeling techniques for stereo SELD tasks, providing a promising direction for future research in audio event and localization analysis.

\bibliographystyle{IEEEtran}
\bibliography{refs}
%
%
%
%
%
%
%
%
%

\end{sloppy}
\end{document}